\documentclass{article}

\usepackage{neurips_2023}

\usepackage[utf8]{inputenc} 
\usepackage[T1]{fontenc}    
\usepackage{hyperref}       
\usepackage{url}            
\usepackage{booktabs}       
\usepackage{amsfonts}       
\usepackage{nicefrac}       
\usepackage{microtype}      
\usepackage{xcolor}         
\usepackage{bbding}
\usepackage{graphicx}
\usepackage{amsmath}
\usepackage{amssymb}
\usepackage{booktabs}
\usepackage{multirow}
\usepackage{enumitem}
\usepackage{bbding}
\setlist{nolistsep}
\usepackage{bm}

\newcommand{\kp}[1]{{\textcolor{blue}{[KP:] #1}}}

\title{Align, Adapt and Inject: Sound-guided\\ Unified Image Generation}

\author{
Yue Yang\textsuperscript{1, 2}
\quad Kaipeng Zhang\textsuperscript{2 \Envelope}
\quad Yuying Ge\textsuperscript{3}
\quad Wenqi Shao\textsuperscript{2}
\\
\quad \textbf{Zeyue Xue\textsuperscript{3}}
\quad \textbf{Yu Qiao\textsuperscript{2}}
\quad \textbf{Ping Luo\textsuperscript{2, 3 \Envelope}}
\\\\
\textsuperscript{1}{ Shanghai Jiao Tong University}
\quad \textsuperscript{2}{Shanghai AI Laboratory} \quad
\textsuperscript{3}{The University of Hong Kong} \quad 
\\
\small{\texttt{yang-yue@stju.edu.cn}
\quad \texttt{\{zhangkaipeng, shaowenqi, qiaoyu\}@pjlab.org.cn}} \\
\small{\texttt{yuyingge@hku.hk} 
\quad \texttt{xuezeyue@connect.hku.hk}
\quad \texttt{pluo@cs.hku.hk}}
}

\begin{document}

\maketitle

\let\thefootnote\relax\footnotetext{\Envelope \, denotes corresponding authors.}

\begin{abstract}
  Text-guided image generation has witnessed unprecedented progress due to the development of diffusion models. Beyond text and image, sound is a vital element within the sphere of human perception, offering vivid representations and naturally coinciding with corresponding scenes. Taking advantage of sound therefore presents a promising avenue for exploration within image generation research. However, the relationship between audio and image supervision remains significantly underdeveloped, and the scarcity of related, high-quality datasets brings further obstacles. In this paper, we propose a unified framework `Align, Adapt, and Inject' (AAI) for sound-guided image generation, editing, and stylization. In particular, our method adapts input sound into a sound token, like an ordinary word, which can plug and play with existing powerful diffusion-based Text-to-Image (T2I) models. Specifically, we first train a multi-modal encoder to align audio representation with the pre-trained textual manifold and visual manifold, respectively. Then, we propose the audio adapter to adapt audio representation into an audio token enriched with specific semantics, which can be injected into a frozen T2I model flexibly. In this way, we are able to extract the dynamic information of varied sounds, while utilizing the formidable capability of existing T2I models to facilitate sound-guided image generation, editing, and stylization in a convenient and cost-effective manner. The experiment results confirm that our proposed AAI outperforms other text and sound-guided state-of-the-art methods. And our aligned multi-modal encoder is also competitive with other approaches in the audio-visual retrieval and audio-text retrieval tasks. 

\end{abstract}

\begin{figure}[htb]
\centering
\scalebox{1}{
\includegraphics[width=1\textwidth]{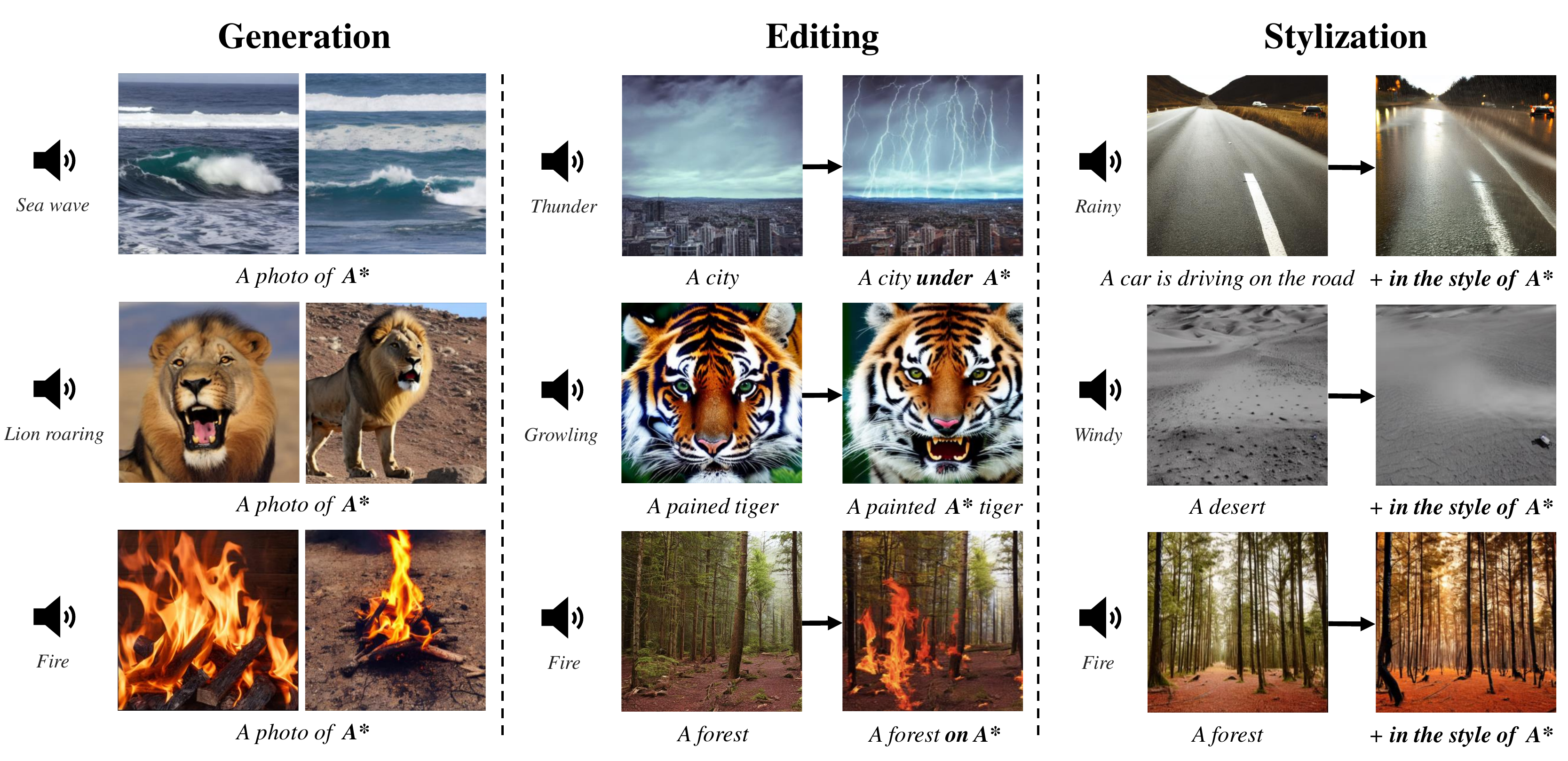}
}
\caption{Examples from our unified strategy AAI. $A^*$ is the audio token for each audio. Our method provides various capabilities based on sound inputs. The user can use the input sound for image generation (left), image editing (middle), and image stylization (right) flexibly.}
\vspace{-4pt}
\label{fig:intro}
\vspace{-2mm}
\end{figure}

\section{Introduction}
\label{headings}

AI Generated-Content (AIGC) research has witnessed rapid advancement, particularly in image generation, image editing, and image stylization \cite{saharia2022photorealistic,ramesh2022hierarchical,rombach2022high,nichol2021glide,yu2022scaling,avrahami2022blended,ho2020denoising,xue2023raphael}. 
In the landscape of generative models, Generative Adversarial Networks (GANs) \cite{goodfellow2020generative} 
have shown phenomenal generative power as a roadmap. Recently, Text-to-Image (T2I) diffusion models have demonstrated an unprecedented ability to outperform GANs. T2I models like Imagen~\cite{saharia2022photorealistic}, DALL-E 2~\cite{ramesh2022hierarchical}, and Stable Diffusion~\cite{rombach2022high}, trained on large text-image pairs, can reason based on human requirements to create fantastic, appealing, and realistic images, which raises infinite possibilities for creative fields. If not specified, T2I models indicate T2I diffusion models in this paper.

Although large-scale T2I models have advanced in leaps and bounds to yield more appealing images, they strained themselves in vision and text modalities, while the discrete form of text makes it challenging to convey distinctions between scenes and vivid properties of objects accurately.
By contrast, 
sound is naturally paired with corresponding scenes, guiding image transformation with vivid signals. For example, thunders with different strengths present varying loudness and characteristic scene, and the tiger performs different expressions when it starts roaring and finishes. But these distinguishing degrees of weather scenes and action changes can only describe by the same single word, ``thunder'' or ``roar''. Meanwhile, as a vital modality for humans to perceive the world, sound provides vivid scenes like weather conditions and animal actions. Therefore, it is promising to generate images with the guidance of sound. 


However, it is challenging to implement an audio-to-image model for sound-guided image generation, like the T2I model. There are mainly two reasons. Firstly, it is impractical to train a large-scale sound-guided generative model like the T2I model, which requires massive manually-annotated audio-image pairs and extremely expensive prohibitive computation costs. Secondly, integrating audio signals into a powerful pre-trained T2I model is also difficult because it needs to align audio representations well with pre-trained visual and text features in the T2I model while the audio-image dataset is small-scale.

To address these issues, recent approaches propose audio representation alignment with the assistance of pre-trained large-scale models. For example, Guzhov \textit{et.al.}~\cite{guzhov2022audioclip} and Lee \textit{et.al.}~\cite{lee2022sound} propose to extend the CLIP \cite{radford2021learning}, a powerful vision-language pretrained model, to audio by introducing an audio encoder and align the audio representation with textual and visual embedding space. Then, Lee \textit{et.al.}~\cite{lee2022sound} further implements sound-guided image editing using the unified embedding and StyleGAN. In addition, GlueGen \cite{qin2023gluegen} attempts to align the representation spaces of audio and text encoders in stable diffusion. Although audio signals are well absorbed into their models, they require training the whole model, incurring a high computation cost. Moreover, there was drastic performance degradation in terms of the quality of generated images when new audio comes. Especially the GlueGen \cite{qin2023gluegen}, which directly feeds the aligned audio embedding into the encoder, output dog-liked animal when input a roaring of tiger in it, and even some indescribable and fusing images under some enviromental sound,  in our comparison experiment. 

\begin{figure}[t]
\centering
\scalebox{1}{
\includegraphics[width=1\textwidth]{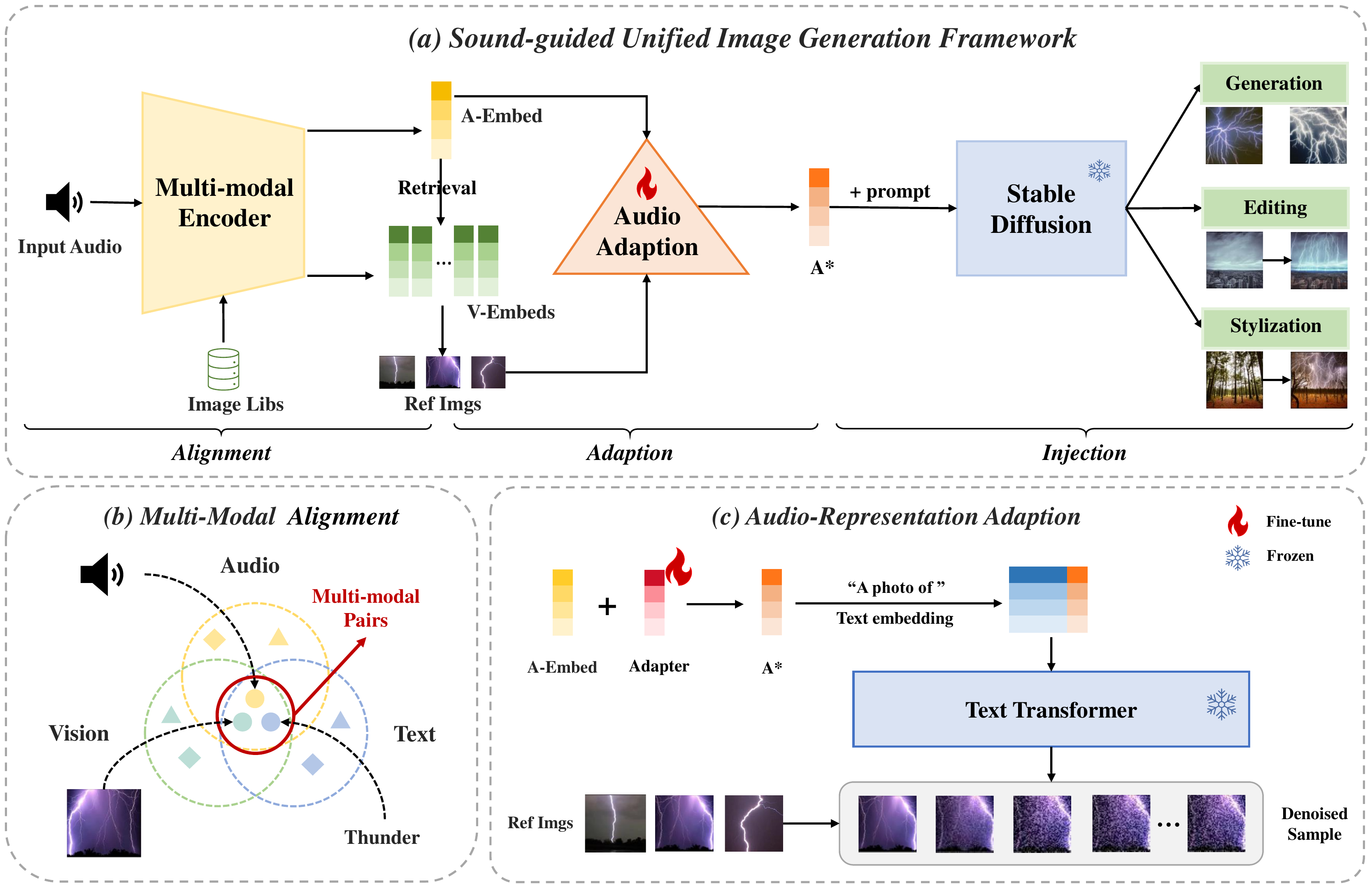}
}
\caption{(a) Overview of Sound-guided Unified Generative Model: Audio representation alignment, adaption, and injection. (b) Contrastive Multi-modal Alignment Stage. (c) Audio-Representation Adaption Stage. There are seriatim elaborate descriptions in Section~\ref{method}.}
\vspace{-4pt}
\label{fig:framework}
\end{figure}

This work proposes a unified strategy called `Align, Adapt, and Inject (AAI)' to accomplish multiple image generation tasks, including image generation, editing, and stylization. While sharing the same north star of leveraging the large-scale T2I models, our model capitalizes on the exclusive benefits of audio for image generation, illustrated in Fig.~\ref{fig:intro}. The AAI comprises three stages, Align, Adapt, and Inject, as depicted in Fig.~\ref{fig:framework}, our method consists of three distinct stages — Align, Adapt, and Inject — aiming to streamline the process of sound-guided image generation in an efficient and effective manner.

In the first stage, we propose a multi-modal alignment module to fine-tune a pre-trained audio encoder. It aligns audio representations with visual and textual modality by training on a small audio-vision-text paired dataset. 
Secondly, to leverage the power of the existing T2I models, we propose an audio adapter that can quickly adapt each audio representation into an audio token under the supervision of a few relevant images. This unique audio token is compatible with T2I models without losing its audible ontological semantics. Finally, we inject the audio token into T2I models plug and play, to complement multi-functional image generation tasks. Therefore, when new audio comes, our strategy can effectively align the audio representation, adapt it into a semantics audio token, and inject it into the T2I models flexibly and efficiently at a low cost.


Our proposed framework is the first attempt to extend sound modality as a token-wise representation into the frozen T2I models plug and play, therefore paving the way for building a unified sound-guided generative model efficiently. In particular, our work makes the following contributions:

\begin{itemize}[noitemsep, topsep=0pt]
    \item We present a sound-guided diffusion models-based image generation framework that simultaneously implements image generation, editing, and stylization.
    \item We train an audio encoder to learn audio representation, which is aligned with the pre-trained textual and visual manifold, and achieves state-of-the-art results in audio-visual retrieval and audio classification benchmarks.
    \item We propose an audio adaptation module that bridges the audio encoder and the frozen T2I model. This module inverts audio representation to a token, similar to an ordinary word, and augments it with relevant images. Thus, the token is compatible with the T2I model and contains vivid and various semantics. As a result, our method can efficiently and effectively take advantage of pre-trained T2I models for sound-guided image generation.
    \item 
    Our experiments demonstrate that our sound-guided generative framework enables more seamless generation and manipulation than based T2I models in various sights, distinct degrees, and higher naturalness. Moreover, compared with the existing sound-guided methods, our method can manage multi-tasks and outperform in many details, like editing local pixels conveniently and generating more obvious and characteristic images for concise scenes or objects. The results are measured in both quantitative and human evaluation.
\end{itemize}

\section{Related Work}
\label{headings}

\textbf{Multi-Modal Alignment.}
In recent years, multi-modal representation aligning \cite{radford2021learning,jia2021scaling,cho2021unifying,wang2021simvlm,chen2022pali} has received unprecedented attention in deep learning, which aims to explore the relationship between different modalities. Benefiting from a large number of vision-language pairs on the Internet, a volume of powerful vision-language foundation models have emerged, like CLIP \cite{radford2021learning} and ALIGN \cite{jia2021scaling}, which demonstrate strong domain adaptation capabilities in vision-language representation. The success of vision-language pre-training inspired the community to shift the emphasis to the audio modality. Guzhov \textit{et al.} \cite{guzhov2022audioclip} proposed AudioClip by inserting an audio encoder into the CLIP-based \cite{radford2021learning} model to align the audio representation with textual and visual embedding space.  The training schema was adopted for the downstream tasks in the latter studies. In work \cite{lee2022sound}, Lee \textit{et al.} leveraged a pre-trained CLIP-based \cite{radford2021learning} model to make audio embedding space suitable for StyleGAN \cite{karras2019style} latent space. However, while these approaches succeeded in representing audio modality, the lack of high-quality paired data for the three modalities caused performance degradation in exploiting the relationships among modalities. To alleviate the problem, some work \cite{gong2022contrastive} began to represent the audio feature in audio-vision or audio-language solely with additional self-supervised methods. 


\textbf{Image Generation, Editing, and Stylization.}
%
Generative Adversarial Nets(GANs) \cite{goodfellow2020generative, abdal2022clip2stylegan,xia2021tedigan,bau2021paint,gal2022stylegan,patashnik2021styleclip,cheng2020rifegan,crowson2022vqgan, jain2022zero,li2019controllable}  have long been the major framework in text-to-image generation. In tandem with CLIP \cite{radford2021learning}, a semantically rich image-text representation model trained on millions of text-image pairs, various GANs-based models \cite{abdal2022clip2stylegan,xia2021tedigan,bau2021paint,gal2022stylegan,patashnik2021styleclip,song2020denoising} achieve remarkable performance. 
Apart from scaling up GANs, Auto-regressive transformer \cite{vaswani2017attention, ramesh2021zero, ding2021cogview} also produced some conspicuous works.
But recently, due to the superior capacity in synthesizing high-quality and diverse images, diffusion models \cite{saharia2022photorealistic, ramesh2022hierarchical, ramesh2021zero, song2019generative, ho2020denoising}  developed rapidly and became the mainstream method in the text-to-image generation field. Large-scale diffusion models, such as Imagen \cite{saharia2022photorealistic}, DALL-E 2 \cite{ramesh2022hierarchical}, and Stable Diffusion \cite{rombach2022high}, have raised the bar for the task on many conditioned text-to-image generation tasks. 
And other volumes of work \cite{avrahami2022blended,hertz2022prompt,kawar2022imagic,ruiz2022dreambooth} began to exploit diffusion models for text-driven image editing. With a small amount of data and without changing the user mode and action, they can transfer different painting or character styles to meet the user's requirements.  
Our main novelties lie in developing a compute-efficient tactic to adapt the audio representation to the powerful diffusion models for multiple image generation.

\textbf{Sound-Guided Image Generation and Manipulation.}
Except for the text modality, audio also rises to prominence as a guiding clue of image generation, which has been proven to represent precise and vivid visual information. Some typical tasks have created models that synthesize images from audio, such as generating talking heads and poses \cite{chen2018lip, zakharov2019few, zhou2019talking, zhou2020makelttalk}. Another new intention is to use audio as a clue to guide image generation. Lee \textit{et al.} \cite{lee2022sound} attempted to manipulate the image to reveal the semantics of the sound by leveraging pre-trained CLIP-based \cite{radford2021learning} audio embedding space. But it can only manipulate the whole image, which cannot apply the sound on the local regions matched with the specific sound. To address this, Li \textit{et al.} showed successful audio-associated image stylization \cite{li2022learning} in an unsupervised manner. Furthermore, as the first attempt to consider audio modality in stable diffusion, GlueGen \cite{qin2023gluegen} leverages parallel corpora to align the representation spaces of some different minority encoders.

\section{Method}
\label{method}

To leverage sound to generate more vivid and realistic images, we propose `Align, Adapt, and Inject'(AAI), a sound-guided image generation framework that can plug and play to pre-trained T2Is and simultaneously implement image generation, editing, and stylization. Our framework bootstraps from large-scale T2Is by adapting the audio representation into an audio token with vivid semantic information, which can be flexibly used in existing powerful T2I models, like an ordinary word. As shown in Fig.~\ref{fig:framework}, it mainly consists of three stages: (i) Multi-modal alignment, (ii) Audio-representation adaption for T2Is, and (iii) Audio-representation injection into T2Is. Section \ref{step1}, Section \ref{step2} and Section \ref{step3} delineates the three stages, respectively.

\subsection{Multi-Modal Alignment}
\label{step1}
In the first stage, we intend to learn audio representation aligned with the paired visual and textual content. To this end, we adopt the pre-trained audio encoder proposed in \textit{CAV-MAE} \cite{gong2022contrastive}, which is trained on two modalities (audio-image). However, considering that only the audio-visual alignment may not be sufficient to utilize the subsequent T2I models, we extend CAV-MAE to CAVT-MAE including three modalities, \textit{i.e.} audio, image, and text, 
 as shown in Fig.~\ref{fig:framework}(b).

\textbf{Multi-modal Encoders.}
To leverage the representation ability of the pre-trained CAV-MAE, we utilize the same audio encoder and vision encoder of it. As for the text modality, We use a text encoder in \cite{rombach2022high} to encode the label of each audio. Although the text encoder does not align with the image encoder in CAV-MAE, we find it the encoded label embedding works well in our CAVT-MAE. 
When fine-tuning CAVT-MAE, we keep both the visual encoder and textual encoder weights frozen and only tune the audio encoder to improve training efficiency and data efficiency. 

\textbf{Loss functions for Aligning Multi-modal Representations.} Our CAVT-MAE is fine-tuned by the alignment between audio and text, audio and vision. The total loss $\mathcal{L}$ is given by: 
\begin{equation}
     \mathcal{L} = \mathcal{L}_c(A,T) + (1-\alpha)\mathcal{L}_c(A,V) + \frac{\alpha}{2} \mathcal{L}_t(A,V),
\end{equation}
where $A, T$, and $V$ denote modalities for audio, text, and vision, respectively. $L_c(\cdot,\cdot)$ is the InfoNCE loss between two modalities, which can align multi-modal representations. $L_t(A,V)$ is a momentum contrastive loss to align audio and visual representation. It utilizes a momentum visual encoder to generate visual representations robust to noise. $\alpha$ is a hyper-parameter. We present two types of losses in the following.

\textbf{InfoNCE Loss \bm{$L_c(A,T/V)$}.} The InfoNCE loss \cite{radford2021learning} is employed to train our audio encoder by pulling the positive pairs together and pushing the negative pairs apart. 
For simplicity, we use $a$ and $m$ to denote an audio and multi-modal sample, respectively, where $m$ can be $v$ and $t$ for image and text, respectively.
Formally, given the $i$-th audio sample $a_i$ in a mini-batch $N$ and its positive paired data $m^+$ and N negative paired data $m^-_j$, where $j\in \{1,2,..., N\}$,
the audio encoder $F_a(\cdot)$ extracts the audio feature $f_a$, and the other modality is mapped to $f_m$ by its specific encoder $F_m(\cdot)$.
The InfoNCE loss gives the similarity of positive pairs from audio to the other modality by: 
\begin{equation}\label{eq:infonce-a2m}
p_i^{\mathrm{a} 2 \mathrm{m}}=\frac{\exp(\operatorname{sim}(f_{a_i}, f_{m^+})/ \tau)}{\sum_{j=1}^N \exp \left(\operatorname{sim}(f_{a_i}, f_{m_j^-}) / \tau\right)}, 
\end{equation}
where $sim$ is the cosine similarity function, and $\tau$ is the temperature. Following the same notation, the similarity of positive pairs from the other modality to the audio is given by: 
\begin{equation}\label{eq:infonce-m2a}
p_i^{\mathrm{m} 2 \mathrm{a}}=\frac{\exp(\operatorname{sim}(f_{m_i}, f_{a^+})/ \tau)}{\sum_{j=1}^N \exp \left(\operatorname{sim}(f_{m_i}, f_{a_j^-}) / \tau\right)}.
\end{equation}
With Eqn. (\ref{eq:infonce-a2m}) and Eqn. (\ref{eq:infonce-m2a}), the InfoNCE loss can be written as:
\begin{equation}
    \mathcal{L}_c(A,T/V)=\frac{1}{2} \mathbb{E}_{(A, T/V) \sim D}\left[\mathrm{H}\left(\boldsymbol{y}^{\mathrm{a} 2 \mathrm{m}}, \boldsymbol{p}^{\mathrm{a} 2 \mathrm{m}}\right)+\mathrm{H}\left(\boldsymbol{y}^{\mathrm{m} 2 \mathrm{a}}, \boldsymbol{p}^{\mathrm{m} 2 \mathrm{a}}\right)\right],
\end{equation}
where $\boldsymbol{p}^{\mathrm{a} 2 \mathrm{m}}$ and $\boldsymbol{p}^{\mathrm{m} 2 \mathrm{a}}$ are vector-forms of $p_i^{\mathrm{a} 2 \mathrm{m}}$ and $p_i^{\mathrm{m} 2 \mathrm{a}}$, respectively. $\boldsymbol{y}^{\mathrm{a} 2 \mathrm{m}}$ and $\boldsymbol{y}^{\mathrm{m} 2 \mathrm{a}}$ are the ground-truth one-hot similarity, where the probability of negative pairs is 0 and the probability of positive pairs is 1.

\textbf{Momentum Contrastive Loss \bm{$\mathcal{L}_t(A,V)$}.} 
For the audio-vision part, the positive paired images are frames extracted from the corresponding video, while the negative ones are from other videos. However, since the audio-video pairs used for fine-tuning are mainly collected from the web, the collected data tend to be noisy, \textit{i.e.} negative videos for the audio may also match the audio's content in contrastive learning. For example, the rainy scene may be accompanied by the sound of wind and thunder. 
%
To alleviate this, we leverage the pseudo-target generated by the momentum visual encoder \cite{li2021align}. It is an exponential-moving-average version and serves as a perpetually evolving teacher.
With the momentum encoder $F^\prime_v(\cdot)$, the similarity between audio-image and image-audio pairs become $\operatorname{sim}(f_a, f^{\prime}_v)$ and $\operatorname{sim}(f^{\prime}_v, f_a)$. The pseudo targets $\boldsymbol{q}^{\mathrm{a} 2 \mathrm{v}}$ and $\boldsymbol{q}^{\mathrm{v} 2 \mathrm{a}}$ are then obtained with similarity operation in Eqn. (\ref{eq:infonce-a2m}) and Eqn. (\ref{eq:infonce-m2a}) replaced. In this way, we can use KL-divergence to construct $\mathcal{L}_t(A,V)$, as given by: 
\begin{equation}
\mathcal{L}_t(A,V)=\mathbb{E}_{(A, V) \sim D}\left[\mathrm{KL}\left(\boldsymbol{q}^{\mathrm{a} 2 \mathrm{v}} \| \boldsymbol{p}^{\mathrm{a} 2 \mathrm{v}}\right)+\mathrm{KL}\left(\boldsymbol{q}^{\mathrm{v} 2 \mathrm{a}} \| \boldsymbol{p}^{\mathrm{v} 2 \mathrm{a}}\right)\right].
\end{equation}

\subsection{Audio-Representation Adaption}
\label{step2}

We aim to implement a sound-guided generative framework by leveraging the power of the existing T2Is. Ideally, the rich semantics in the audio should be preserved, and the audio modality can guide the visual transformation intuitively like the text modality. Inspired by \cite{gal2022image}, where the visual concepts were inverted into a continuous representation, we introduce an audio adapter to transform audio representations into an intermediate form for T2Is.

In this stage, our goal is to find a new representation $A^*$ which can be used for various image-generative tasks.
Following the method in \cite{gal2022image}, we adopt the \textit{visual reconstruction} as the adaption object. As shown in Fig.~\ref{fig:framework}(c), we outline the details of the adaptation process. In the process, we minimize the LDM loss with a trainable adapter, $f_{Adapter}$, while freezing other parameters. The initial $f_{Adapter}$ is a series of randomly sampled neutral context parameters, which is the same sized as the aligned audio representation. The LDM loss is introduced in Latent Diffusion Model (LDM) \cite{rombach2022high}. Its optimization object is to denoise a noisy latent representation and can be defined as:
$
\mathbb{E}_{z, y, t, \epsilon \sim \mathcal{N}(0,1)}\left\lfloor\left\|\epsilon-\epsilon_\theta\left(z_t, t, c(y)\right)\right\|^2\right\rfloor,
$
where $t$ represents the timestep, $z_t$ denotes the latent noisy representation at timestep $t$, $\epsilon$ is an unscaled noise sample, and $\epsilon_\theta$ is the denoising network based on a U-Net.
In our work, $y$ is the text prompt and $c(\cdot)$ is the text encoder implemented by a Text Transfomer \cite{vaswani2017attention}. Then the loss of the audio-representation adaption can be formulated as follows:
\begin{equation}
\mathcal{L}_{A}=\mathbb{E}_{z, y, f_a, t, \epsilon \sim \mathcal{N}(0,1)}\left\lfloor\left\|\epsilon-\epsilon_\theta\left(z_t, t, c([y, f_a + f_{Adapter}])\right)\right\|^2\right\rfloor,
\end{equation}
where the prompt is ``a photo of *'', $[\cdot, \cdot]$ means the concatenate operation, and * denotes the token representing $A^*$. We expect that the learnable $f_{Adapter}$ can assist in aligning the audio representation $f_a$ within the text embedding space by directly optimizing the aforementioned loss function.
To train the $f_{Adapter}$, we use a small set of reference images ($3$\textasciitilde$5$) retrieved by the audio representation from the Image library. Importantly, after obtaining the $f_{Adapter}$, we inverse it to a special pseudo-token $A^*$ for the audio-representation injection.

\subsection{Audio-Representation Injection}\label{step3}

In this stage, our objective is to integrate the newly introduced pseudo-token $A^*$ into the T2Is framework to enable image generation, editing, and stylization. By treating the audio adapter token $A^*$ as a regular word, we can seamlessly incorporate it into coherent text sentences and leverage it in our innovative unified sound-guided generative model. When adopted for image generation, each $A^*$ is incorporated into new conditioning to compose novel scenes by texts. In the context of editing and stylization, users have the flexibility to introduce new tokens or replace existing ones in the prompts. Following \cite{hertz2022prompt}, we incorporate the attention maps of the source image based on the modified prompt's attention map $A^*$ to ensure the preservation of the source image's composition.

\section{Experiments}
\label{headings}

\subsection{Experiments Setup}
\textbf{Datasets.}
Following previous experimental protocols, we use a publicly available dataset \textit{VGG-Sound} \cite{chen2020vggsound} in our studies. \textit{VGG-Sound} contains large-scale audio-visual pairs over 200K. Due to missing video clips on \textit{YouTube}, we collect 180K videos in total and follow the widely-used splits to divide the dataset into 166K and 14K for training and testing, respectively. 

\textbf{Implementation Details.}
All of our experiments are performed on NVIDIA A100
GPUs with PyTorch framework \cite{paszke2017automatic}. In the fine-tuning stage, we adopt the pre-trained models in CAV-MAE \cite{gong2022contrastive} and extend the model by adding text modality. For the text encoder, we employ a Bert-like \cite{devlin2018bert} model as the text encoder used in the Stable Diffusion Model \cite{rombach2022high}. All of the above encoders have the same output dimensions, 768, and the parameters of the vision and text encoder are frozen during fine-tuning.
We use the mini-batch AdamW optimizer \cite{loshchilov2017decoupled} with a weight decay of 0.02. 
For the momentum updating, we set $m=0.995$. Other settings are the same with \cite{gong2022contrastive}. And to adapt the audio inputs into the T2I models, we apply the Stable Diffusion Model (SDM) \cite{rombach2022high} as the generation baseline and retain the original hyper-parameter choices. 



\subsection{Applications}
Our method, described in Section \ref{method}, enables image generation, editing, and stylization by adapting the audio to T2Is. In the following sections, we show a range of applications using our framework.

\begin{figure}[htb]
\centering
\scalebox{0.98}{
\includegraphics[width=1\textwidth]{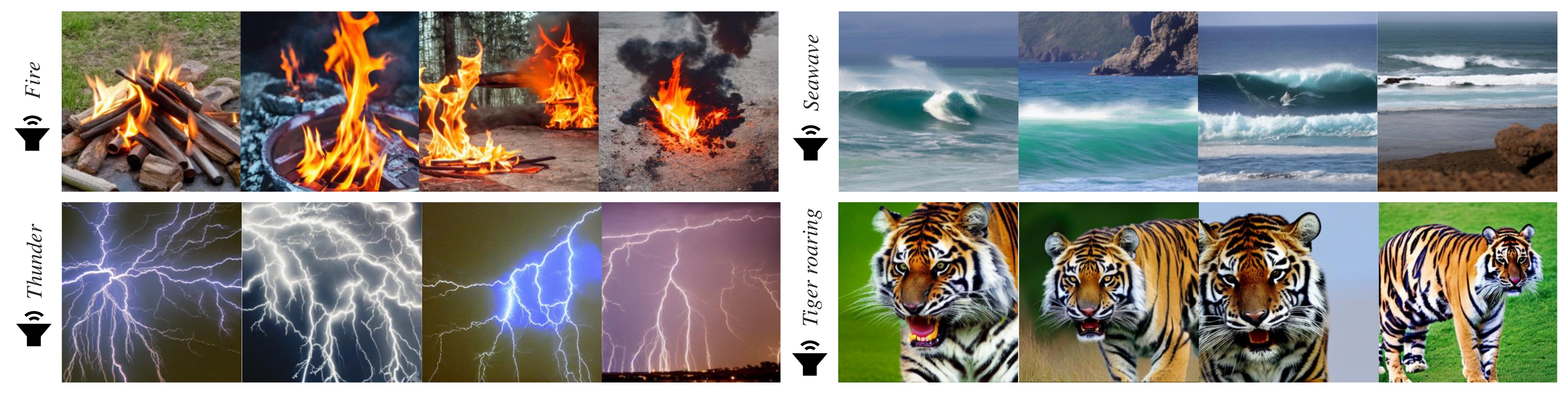}
}
\caption{Results of the image generation given different audio inputs.}
\vspace{-1mm}
\label{fig:generation}
\end{figure}

\textbf{Image Generation.}
We start by demonstrating the ability in image generation. Each sound token $A^*$ inverted from the sound adapter is served as a high-level semantic of a scene or object expressed by sound. We use stereotyped prompts to generate the audible object, like ``A photo of $A^*$'', ``A painting of $A^*$'', ``A sight of $A^*$'' \textit{etc.} and input them to the T2I model. 
We depict some examples in Fig.~\ref{fig:generation}. As the results demonstrate, our model can effectively capture the semantics in the audio and provides sufficient details, such as the variation of scenes (\textit{e.g.} thunder, fire, sea wave) and the actions of objects (\textit{e.g.} tiger growling). And the generated images are greatly diverse in shapes and colors. For example, thunders sometimes appear loud with a criss-cross shape and sometimes appear soft with a small range of strips, these can be reflected in the samples of thunder in Fig.~\ref{fig:generation}. 


\begin{figure}[htb]
\centering
\scalebox{0.98}{
\includegraphics[width=1\textwidth]{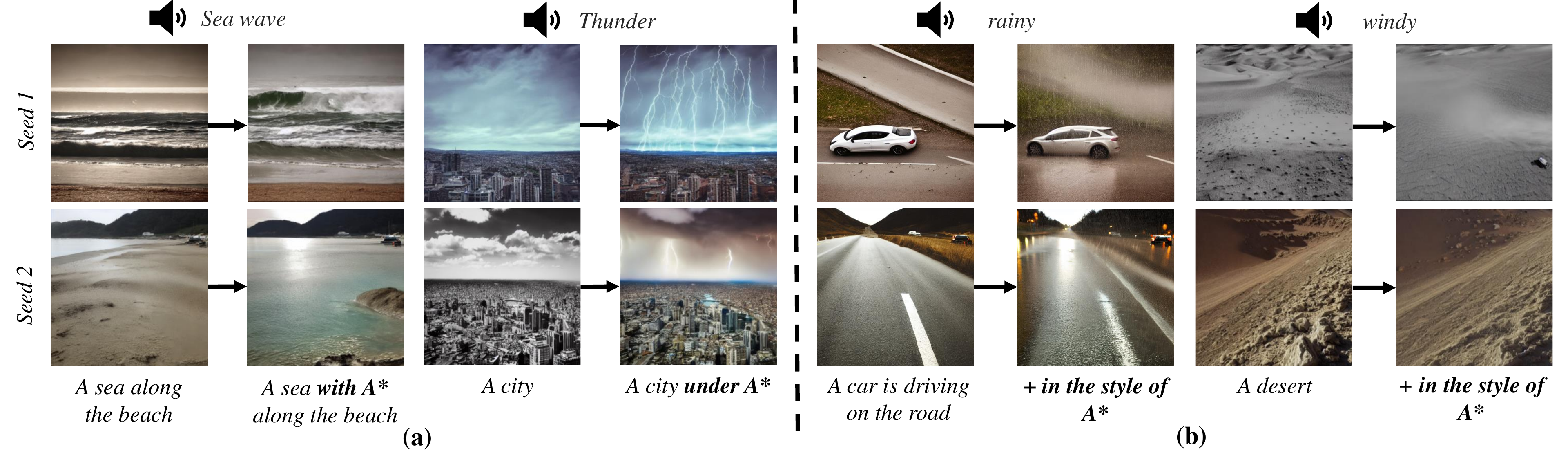}
}
\caption{Results of the image editing (a) and stylization (b) given different audio inputs.}
\vspace{-10pt}
\label{fig:manipulation}
\end{figure}

\paragraph{Image Editing.}
In this section, we show our model's preliminary sound-guided image editing ability. 
We employ some pairs of source prompts and target prompts to manipulate the local pixel of the image, which corresponds and is supervenient to the audio. As shown in Fig.~\ref{fig:manipulation}(a), we inject the $A^*$ into source prompts like ``A sea along the beach'', ``A city'', therefore they turned into ``A city under $A^*$'', ``A sea with $A^*$ along the beach''. It is obvious that the state of the sea is changed and the thunder is added above the city. It's worth noting that our model can preserve the image composition completely, which is valuable for editing. Our model also edits the image following audio semantics, such as changing the state of the sea and adding the thunder above the city.

\paragraph{Image Stylization.}
A typical essential characteristic is that different weather is usually accompanied by different sounds. Therefore, various properties can be caught by the sound of matched weather. Here, we show that our AAI can also use the audio with the adapter to represent a specific, vivid style. To blend such audio, we add it on the end of any ordinary prompt with a phrase ``in the style of $A^*$'', like `` a forest in the style of $A^*$''. As Fig.~\ref{fig:manipulation}(c) shows, by adding the $A^*$ style, the car is turned driving in the misty rain, and the forest is tinged with fiery red. These various visual scenes are captured by our sound adapter successfully. 


\begin{table}[htb]
\footnotesize
\centering
\caption{Audio-Video retrieval results on VGGSound. w/o $m$: fine-tuning without the pesudo-targets}\label{tab:retrieval}
\resizebox{\linewidth}{!}{
\begin{tabular}{@{}lcccccccccc@{}}
\toprule
& \multirow{2}{*}{Fine-tune} & \multicolumn{3}{c}{EvalSubset-1.4K} & \multicolumn{3}{c}{EvalSubset-2.8K} & \multicolumn{3}{c}{EvalSubset-14K}  \\ \cmidrule(l){3-11} 
&  & R@1  & R@5  & R@10   & R@1  & R@5  & R@10   & R@1  & R@5  & R@10 \\ \midrule

CAV-MAE \cite{gong2022contrastive} & -                         & 0.28     & 0.77     & 1.62   & 0.17   & 0.60   & 1.41                & 0.02    & 0.15    & 0.24  \\

AudioCLIP \cite{guzhov2022audioclip}             & -                         & 13.84  & 33.26  & 44.42  & 8.36   & 23.79   & 34.28                & 2.69    & 9.02    & 14.15 \\
AudioCLIP             & \checkmark         & 21.53  & 47.24  & 57.98                   & 14.68   & 36.82   & 48.34                & 5.67    & 17.17    & 25.34     \\

CAVT-MAE w/o $m$                  & \checkmark                       & 27.04  & 53.31  & 64.75  & 20.76   & 44.87   & 55.47                & 7.96 & 22.26 &  31.64               \\
CAVT-MAE\textsuperscript{Ours}      & \checkmark       & \textbf{28.31}  & \textbf{54.51}  & \textbf{64.26}   & \textbf{21.53}   & \textbf{45.05}   & \textbf{55.89}       & \textbf{8.86} & \textbf{23.46} & \textbf{32.83}                  \\ \bottomrule
\end{tabular}
}
\end{table}

\begin{table}[h]
\centering
\begin{minipage}[c]{0.52\linewidth}
\caption{Zero-Shot audio classification results. Comparisons with existing multi-modal methods on the ESC-50 and the Urban sound 8k, such as AudioClip ~\cite{guzhov2022audioclip}, Wav2clip~\cite{wu2022wav2clip}, and SGSIM~\cite{lee2022sound}. Modal: Data modalities. $*$: Fine-tuned on the VGG-Sound.} \label{tab:zero-shot}
\resizebox{\linewidth}{!}
{
    \begin{tabular}{@{}lccc@{}}
        \toprule
        \multirow{2}{*}{Model} & \multirow{2}{*}{Modal} & \multicolumn{2}{c}{Dataset} \\ \cmidrule{3-4}
        & & ESC-50 & Urban sound 8k \\ \midrule
        AudioCLIP~\cite{guzhov2022audioclip} & A-V-T & 4.1 \% & 18.1 \% \\
        AudioCLIP$^*$~\cite{guzhov2022audioclip} & A-V-T  & 6.2 \% & 19.58 \% \\
        Wav2clip~\cite{wu2022wav2clip} & A-V  & 41.4 \% & 40.4 \% \\
        SGSIM~\cite{lee2022sound} & A-T  & \textbf{57.8 \%} & \textbf{45.7\%} \\
        CAVT-MAE\textsuperscript{Our} & A-V-T  & \underline{42.8 \%} & \underline{39.8 \%} \\
        \bottomrule
    \end{tabular}
}
\end{minipage}\hfill
\begin{minipage}[c]{0.46\linewidth}
\caption{Audio-Text retrieval results on \textit{VGG-Sound}, w/o $m$: fine-tuning without the pesudo-targets.}\label{tab:cls}
\resizebox{\linewidth}{!}{
    \begin{tabular}{@{}lcc@{}}
        \toprule
        Model & Fine-tune & Accuracy \\  \midrule
        AudioCLIP~\cite{guzhov2022audioclip} & - & 38.94\%  \\ 
        AudioCLIP~\cite{guzhov2022audioclip} & \checkmark & 48.41\% \\
        \midrule
        CAVT-MAE w/o $m$ & \checkmark & 49.32\%  \\
        CAVT-MAE\textsuperscript{Our} & \checkmark & \textbf{52.16\%} \\
        \bottomrule
    \end{tabular}
}
\end{minipage}
\end{table}

\subsection{Quantitative Results}
\textbf{Multi-Modal Representation Alignment.}
To gauge the generalization of the aligned audio representations, we compare our method with others \cite{guzhov2022audioclip,wu2022wav2clip,lee2022sound} on downstream tasks, including audio-visual retrieval and audio classification method. We evaluate our model by benchmarking the audio-visual retrieval tasks on different test subsets of \textit{VGG-Sound} by following the standard evaluation protocol. As shown in Table~\ref{tab:retrieval},  with fine-tuning, our approach achieves the best performance. Compared with the fourth line in Table~\ref{tab:retrieval}, directly finetuned by InfoNCE \cite{radford2021learning}, we improve the audio representation alignment by the pseudo-targets. Moreover, we perform the zero-shot transfer to the audio classification and compare it with the existing methods on the ESC-50 \cite{piczak2015esc}
and the Urban sound 8k \cite{salamon2014dataset}. Table~\ref{tab:zero-shot} shows that our model can achieve remarkable results compared with the models pre-trained on the dataset with three modalities. To further investigate
the alignment between audio and text, we also conduct audio-text retrieval experiments on \textit{VGG-Sound}, as shown in Table~\ref{tab:cls}.

\begin{figure}[htb]
\centering
\scalebox{1}{
\includegraphics[width=1\textwidth]{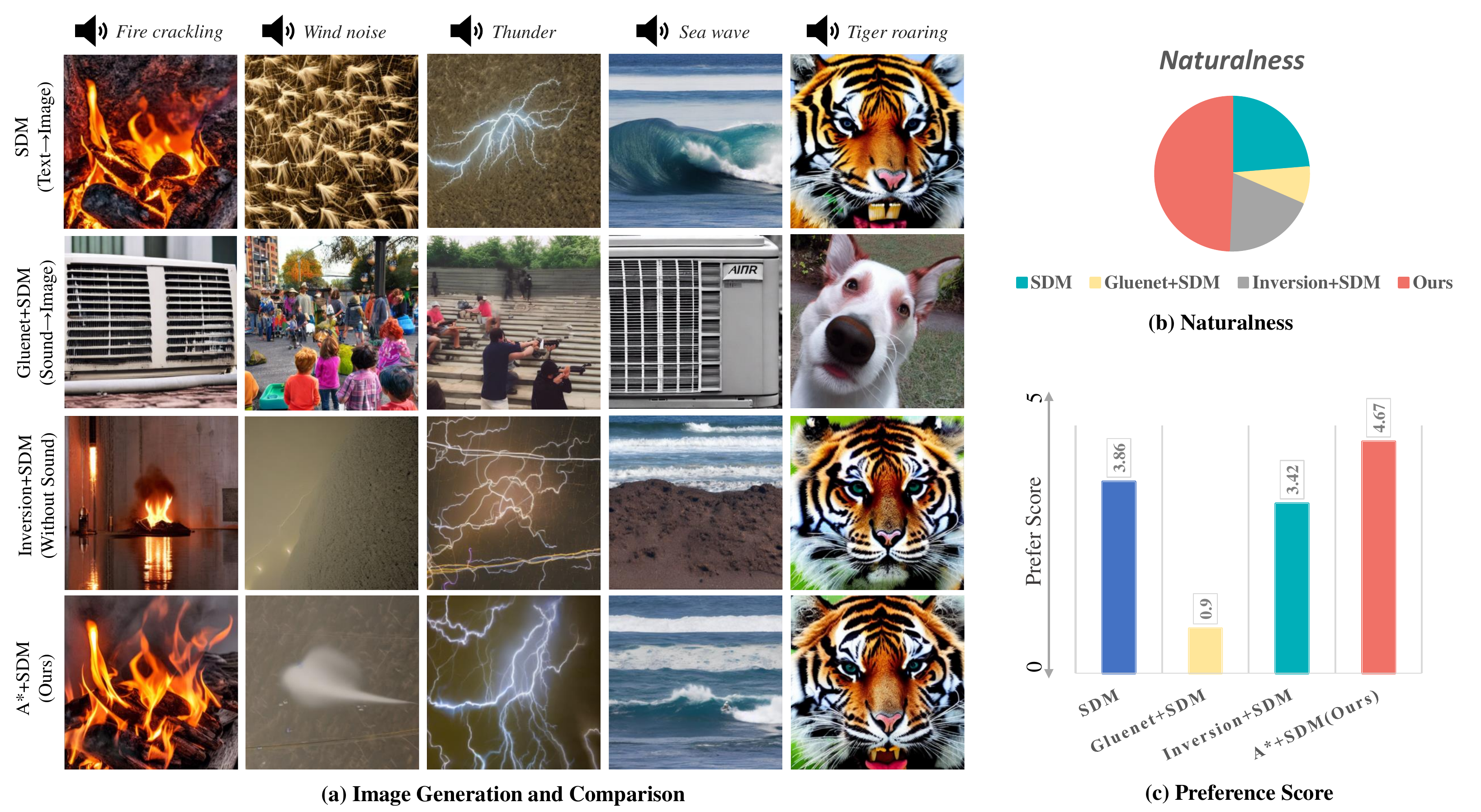}
}
\caption{Qualitative comparisons to alternative personalized creation approaches. (a) Our model can more accurately capture the semantics of the audio, enabling generate images which typically more faithful images to the input. (b) Human evaluation results comparing Ours vs. other existing models.}
\vspace{-4pt}
\label{fig:comparsion}
\end{figure}

\textbf{Visual Comparisons.}
In Fig.~\ref{fig:comparsion}, we provide visual comparisons of the different methods. By comparing with SDM \cite{rombach2022high}, GlueNet \cite{qin2023gluegen}, and Textual Inversion \cite{gal2022image}, we can conclude that our sounded-guided approach achieves a better quality of the generated images where existing sound-guided models \cite{qin2023gluegen} would be challenging to preserve the audio identity and leveraging it for the generation. As Fig.~\ref{fig:comparsion} shows,
text-guided models \cite{rombach2022high, gal2022image} are able to generate appealing images. Nevertheless, it continues to encounter challenges with only capturing the most prominent features but failing to provide sufficient details for generating finer features, as mentioned in \cite{gal2022image}. Our model can capture the vivid semantic information of the given audio input and show more radical results, like the tiger growling more fiercely. Moreover, we demonstrate that our proposed method can transfer a more specific and intense style into source images and generate high-quality images since each audio sample has its own context, which makes the guidance richer than the text.

\subsection{User Study}
To further evaluate the effectiveness of our models, we hold a user study. Since we leverage T2I models for our framework, we conduct human evaluations on the original SDM~\cite{rombach2022high}, Image Inversion~\cite{gal2022image}, and the SDM-based sound-guided image generation model GlueNet~\cite{qin2023gluegen}. For a fair comparison, we compare them in the Image generation tasks. Following the previous work~\cite{lee2022sound,lee2022lisa}, we recruited 50 participants and created two questionnaires, including naturalness and preference scores. First, users were asked to assess the naturalness of the images. The score ranges from 1 (low naturalness) to 5 (high naturalness). In the second questionnaire, users were required to assign a preferred score to each image provided with the above models. We report the results in Fig.~\ref{fig:comparsion}(b) and (c). It shows that our method excels in other models in terms of naturalness and user preference.


\subsection{Ablation Study}
\textbf{The Number of Reference Images and Optimization Steps.}
We evaluate the impact of the number of reference images and optimization steps in the audio adaption process. We note that the adapter can capture different semantics and facilitate the T2I model to generate various with different numbers of reference images and optimization steps. As it can be seen in the \textcolor{blue}{supplementary materials}, better results can be achieved with fewer steps and reference images in some cases. 


\begin{figure}[htb]
\centering
\scalebox{0.9}{
\includegraphics[width=1\textwidth]{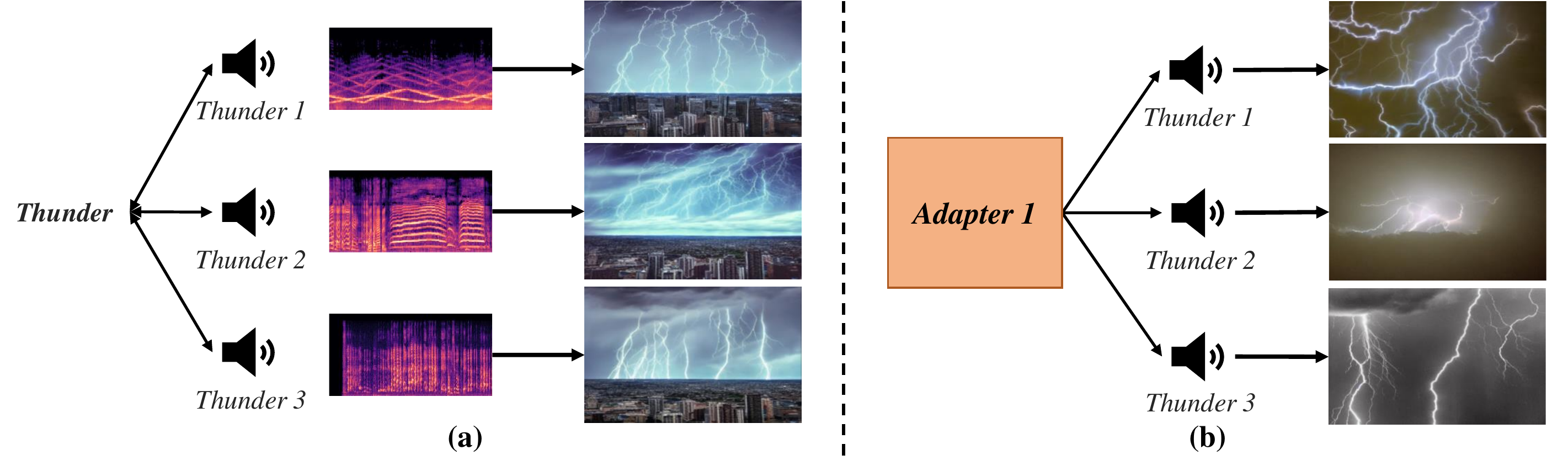}
}
\caption{(a) Qualitative analysis for various reference images in the same kind of sound. (b) Additional results using one adapter with different audio embeddings to the T2I model.}
\vspace{-4pt}
\label{fig:ab2}
\end{figure}

\textbf{The Adaptation Ability on Various Reference images}
We further explore the role of reference images in the audio adaption process. Specifically, we adapt the audio with different reference images from the same class. As Fig.~\ref{fig:ab2}(a) shows, different reference images bring various semantics to the audio adapter and help generate more appealing images. What's more, with the analysis of these images, we find that the prominent features of the images are retained, and fine details would be changed by the effect of different reference images. In the further experiment, there is a surprising finding that one audio adapter could adapt the other audio from the same class to the T2I model. As shown in Fig.~\ref{fig:ab2}(b), it also proves the effectiveness of our audio representation alignment.

\textbf{Analysis on Audio Adapter.}
As depicted in Fig.~\ref{fig:comparsion}, the fourth line represents images generated by our model, which are tailored to each distinct audio input. In contrast, the third line corresponds to the relative version \cite{rombach2022high, gal2022image}, where we eliminate the audio adapter while keep all other settings unchanged, so it only uses reference images to invert the vision-based representation. It is evident that our model adeptly captures the dynamic semantic information conveyed by the audio input, resulting in more extreme outcomes, such as a more fiercely growling tiger. Without the audio adapter, despite maintaining all other settings, the generated images still struggle to capture more than the most salient features, lacking the detail necessary for generating finer attributes. This limitation arises primarily because each audio sample possesses its unique context, thus offering guidance that is richer than a vision-based representation alone.

\section{Conclusion and Limitations}
\label{headings}

\textbf{Conclusion.}
In this work, we present a unified framework for sound-guided image generation, editing, and stylization. From the external knowledge view, this work is the first step towards implementing a sounded-guided generative model for multi-tasks. In the framework, our model aligns the representation of the user-provided audio and adapts it into an audio token, which has vivid semantics and can be injected into existing T2I models flexibly. We show that our method can leverage the powerful capabilities of T2I models effectively and efficiently without losing audio semantics via the proposed audio adapter. Realistic and vivid results demonstrate that our method successfully captures and utilizes the semantic cue of the audio for generative tasks. 


\textbf{Limitations.}
Our framework is still subject to a limitation that the global audio representation adopted currently bounds the ability to perform more precise and detailed sound-guided generation. The injection of more fine-grained representation will be thoroughly studied in the future. 

\bibliographystyle{unsrt}
\bibliography{reference}

\clearpage
\appendix

\section{Detailed Method}
\label{method}
As delineated in the primary body section, our ``Align, Adapt and Inject'' (AAI) model comprises three stages: Alignment, Adaptation, and Injection. The Alignment stage was comprehensively discussed in Section 3.1. In the following passage, we will provide a detailed explanation of the methodologies employed in the Injection stage of our framework.

\subsection{Audio-Representation Injection into T2Is.}
\label{step3}
With the audio-adapted token $A^*$, we intend to treat it as a word, which can be seamlessly inserted into a coherent text sentence and then used for our novel, unified, sound-guided generative framework. Notably, this process leaves the generative model untouched and frozen. In this stage, our goal is to inject this new pseudo-token $A^*$ into the T2Is to achieve sound-guided image generation, editing, and stylization. When adopted for image generation, each $A^*$ is incorporated into specific conditioning to compose novel scenes by texts. While for editing and stylization, following \cite{hertz2022prompt}, we inject the attention maps of the source image with the modified prompt $A*$.


\subsubsection{Image Generation}
For image generation, $A^*$, along with its accompanying audio representation, serves as a high-level semantic depiction of an object or scene expressed by sound. Therefore, we use some stereotyped prompts to generate the audible object, like ``A photo of $A^*$'', ``A painting of $A^*$'', ``A sight of $A^*$'' \textit{etc.}. In this way, the frozen T2I models can jointly reason over the adapted audio representation and its large body of prior knowledge, bringing them together in a newly-generated image.

\subsubsection{Image Editing}
From \cite{hertz2022prompt}, we are cognizant that the structure and appearances of the generated image depend not only on the random seed but also on the interaction between the pixels to the text embedding through the diffusion process in T2Is. And the main challenge of sound-guided image editing is to preserve the original composition while simultaneously signifying the content embedded in the audio. To this end, we follow \cite{hertz2022prompt} to inject the attention maps of the source image into the generation with the modified prompt with $A*$. 

For image editing, the user can add new tokens or replace some tokens in the prompts. Formally, we can denote the editing process as follows:
\begin{equation}
   \left(\operatorname{Sound-Guided Edit}\left(S_t, S_t^*, t\right)\right)_{i, j}:= \begin{cases} \left(S^*_t\right)_{i, j} & \text{if} \ j= \text{Indexof}(A^{*}) \\ \left(S_t\right)_{i, IndexProj(j)} & \text { otherwise }\end{cases}.
\end{equation}
where $S_t$ and $S^*_t$ are the attention map of the source and modified prompt at time-step $t$ in the diffusion process. $IndexProj(\cdot)$ is a function that accepts the token index in the modified prompt and produces the token index in the source prompt.

\subsubsection{Image Stylization}
Particular sounds may hint human with some specific overall scenes. For example, one of the essential characteristics is that much different weather is usually accompanied by different sounds, like wind and thunder, the sounds we hear often convey the visual textures within a scene. What's more, the style is totally distinguished under specific scenarios. In general, various properties can be caught by the sound of matched weather. Therefore, we show that our unified sound-guided generative model can also use the audio with the adapter to represent a specific, vivid style. To blend such audio, we add it on the end of any ordinary prompt with a phrase ``in the style of $A^*$'', like ``A forest in the style of $A^*$''. With such prompts, the audio-token $A^*$ can take advantage of T2Is and reflect various visual scenes captured by sound under corresponding style conditions. 


\subsubsection{Loudness Reweighting}
In particular, the sound is equipped with loudness inherently, therefore, loudness is a unique and significant attribute of sound, implying various degrees of scenes. According to the loudness of the sound, the user likely manipulates the image by strengthening or weakening the audible content in the image. We can implement it by scaling the weight of the token $A^*$, which can be formulated as:
\begin{equation}
   \left(\operatorname{Sound-Guided Edit}\left(S_t, S_t^*, t\right)\right)_{i, j} = scale \cdot \left(S^*_t\right)_{i, j} \quad if \ j \in ScaleIndex .
\end{equation}
where the $ScaleIndex$ is a set of indexes containing indexes of tokens that need to be scaled, and the $scale$ is the degree of loudness contained in the sound, which is defined in the range of $\left[0,2\right]$. 

\begin{figure}[htb]
\centering
\scalebox{1}{
\includegraphics[width=1\textwidth]{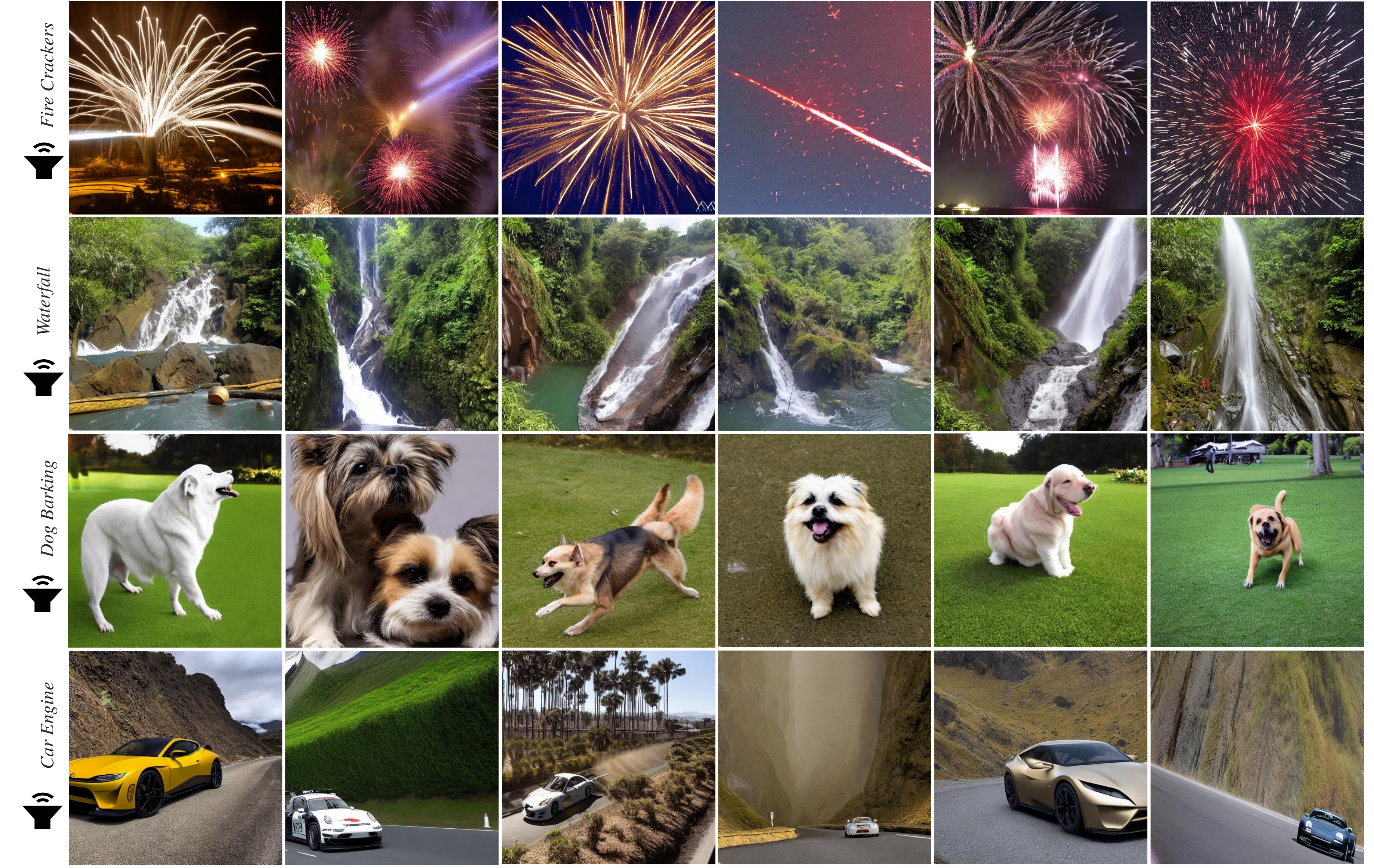}
}
\caption{Generated results of image generation.}
\vspace{-4pt}
\label{fig:gen}
\end{figure}

\section{Experiments}
\label{exp}
\subsection{Data Preparation}
To gain the multi-modal audio representation, we align audio with text and vision modalities. Training a tri-modal contrastive learning model requires datasets paired across audio, text, and visual dimensions. However, the lack of a high-quality audio-visual dataset makes it challenging to secure suitable paired audio-visual inputs. To solve this problem, we follow AudioClip \cite{guzhov2022audioclip} to process the multi-modal dataset. We employ the \textit{VGGSound} dataset for fine-tuning, which comprises 200,000 audio videos. In detail, for the audio modality, we extract audio in the format of \textit{wav} from each video. For the text modality, we take the label of each audio as text input.  
Particularly for the visual modality, given that the majority of videos in \textit{VGGSound} are 10-second clips, we initially extract 20 frames per video. Subsequently, we use CLIP \cite{radford2021learning} as a pre-selector to choose the top 10 frames that most matched with the label from the extracted 20 frames, and we serve the 10 frames as the visual input for each audio.

\subsection{More Results}
\subsubsection{Image Generation}
\label{generation}
To generate images, in Fig~.\ref{fig:gen}, we adapt four different sounds into $A^*$. They are produced by fire-crackers, waterfall, dog-barking, and car-engine, respectively, including voiced scenes like fire-crackers and audible objects like dogs. And then, we regard the $A^*$ as a common word to inject them into T2Is. We send the simple prompt ``A photo of $A^*$'' to Stable diffusion and generate images. In Fig~.\ref{fig:gen}, the shape of our generated fire-crackers is precise and various, embodying very natural effects, just like the real night with fireworks. And the mouse of each generated dog is obviously opening, implying the sound of barking.

\begin{figure}[htb]
\centering
\scalebox{1}{
\includegraphics[width=1\textwidth]{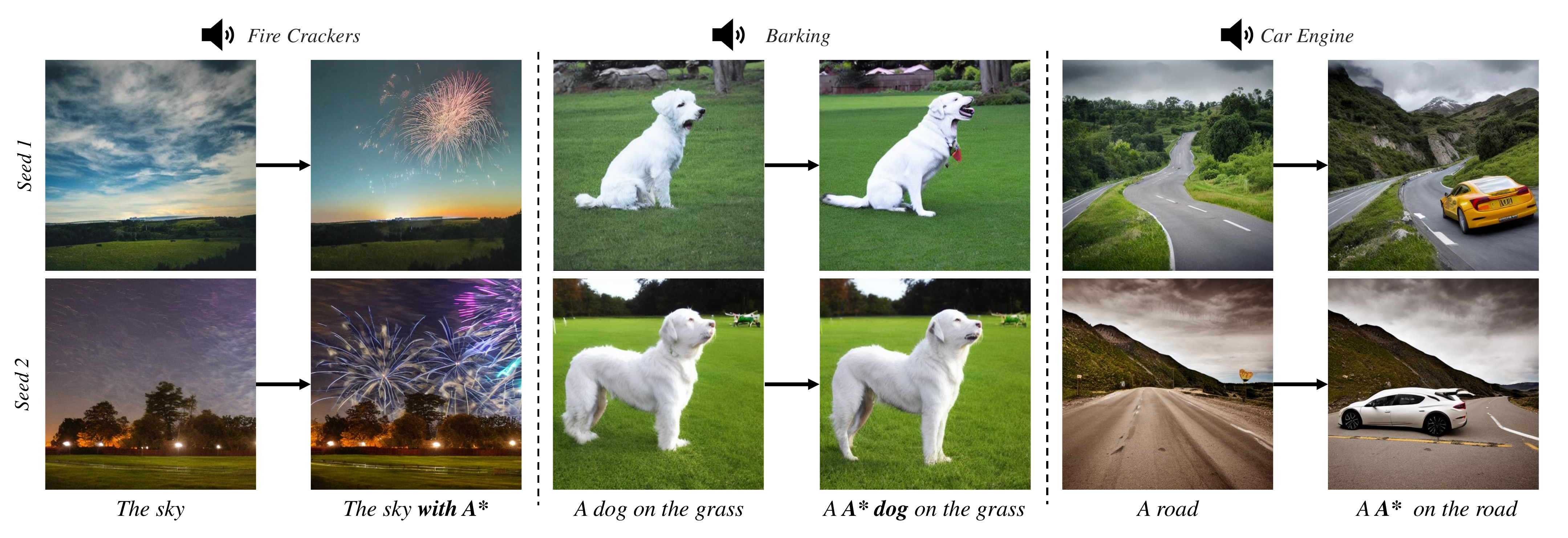}
}
\caption{Generated results in image editing.}
\vspace{-4pt}
\label{fig:edit}
\end{figure}

\subsubsection{Image Editing}
In order to edit images generated by original text prompts, we incorporate $A^*$ into text prompts to facilitate localized editing. Consequently, we employ certain pairs of source and edited prompts for image editing, as shown in Fig.~\ref{fig:edit}. It is obvious that our sound token $A^*$ can act on the specific object within the source images, such as firing crackers in the sky and provoking dogs to bark. It is notable that we only edit the local pixels while remaining the other pixels the same to the maximum extent, which is a troublesome obstacle in the field of sound-guided generation. 

\begin{figure}[htb]
\centering
\scalebox{1}{
\includegraphics[width=1\textwidth]{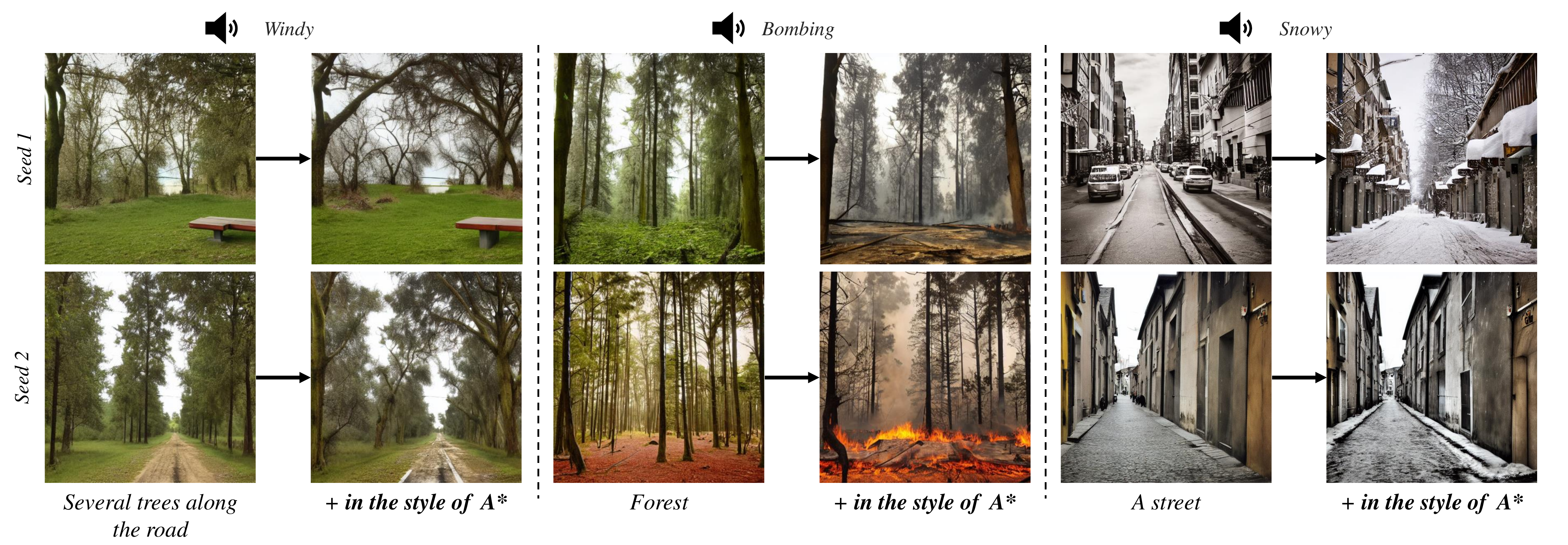}
}
\caption{Generated results of image stylization.}
\vspace{-4pt}
\label{fig:sty}
\end{figure}

\subsubsection{Image Stylization}
As shown in Fig.~\ref{fig:sty}, stylistic transformations of the images are apparent.  In the left column, the trees show a fuzzy and slanted appearance under the influence of the ``windy" style. The middle column displays a forest scene characterized by a smoke-filled and incendiary ambiance, indicative of the ``bombing" style. And the right column presents an entire scene blanketed in snow, thus embodying a ``snow-white" style.

\subsection{Ablation Study}
\subsubsection{The Number of Reference Images and Optimization Steps}
\begin{figure}[htb]
\centering
\scalebox{1}{
\includegraphics[width=1\textwidth]{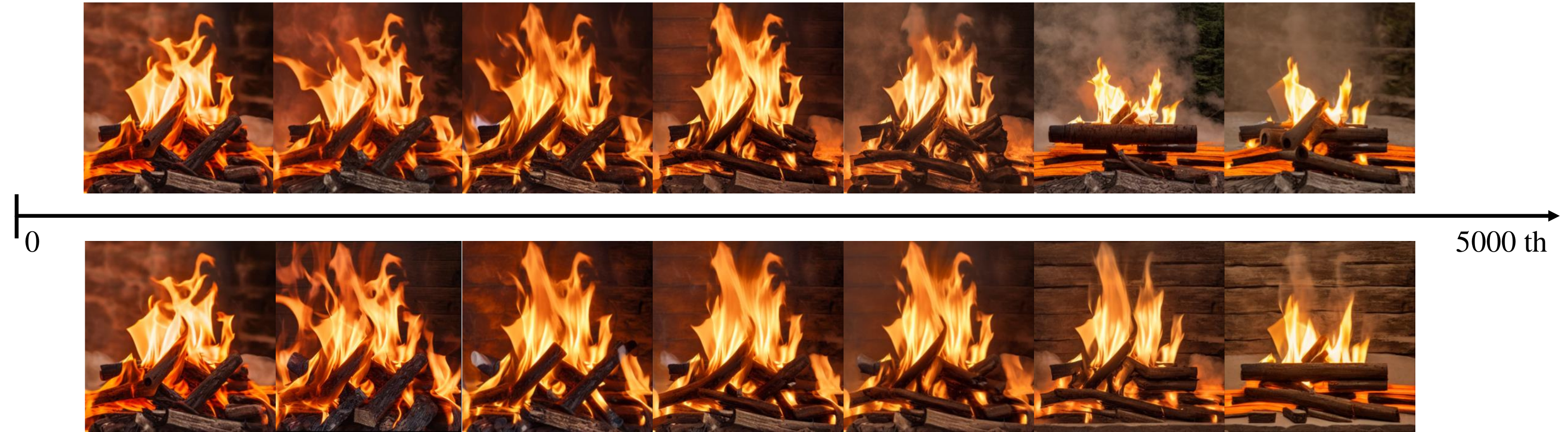}
}
\caption{Ablation study of the adaption process on the number of reference images and optimization steps.}
\vspace{-4pt}
\label{fig:ab1}
\end{figure}

We evaluate the impact of the number of reference images and optimization steps in the audio adaption process. We note that the adapter can capture different semantics and facilitate the T2I model to generate various with different numbers of reference images and optimization steps. 
Fig.~\ref{fig:ab1}, better results can be achieved with fewer steps and reference images in some cases. The results indicate that more reference images would push more constraints in the adaption stage, and more optimization steps contribute to capturing more details.

\subsubsection{Different Loudness of Input Sound}
\begin{figure}[htb]
\centering
\scalebox{1}{
\includegraphics[width=1\textwidth]{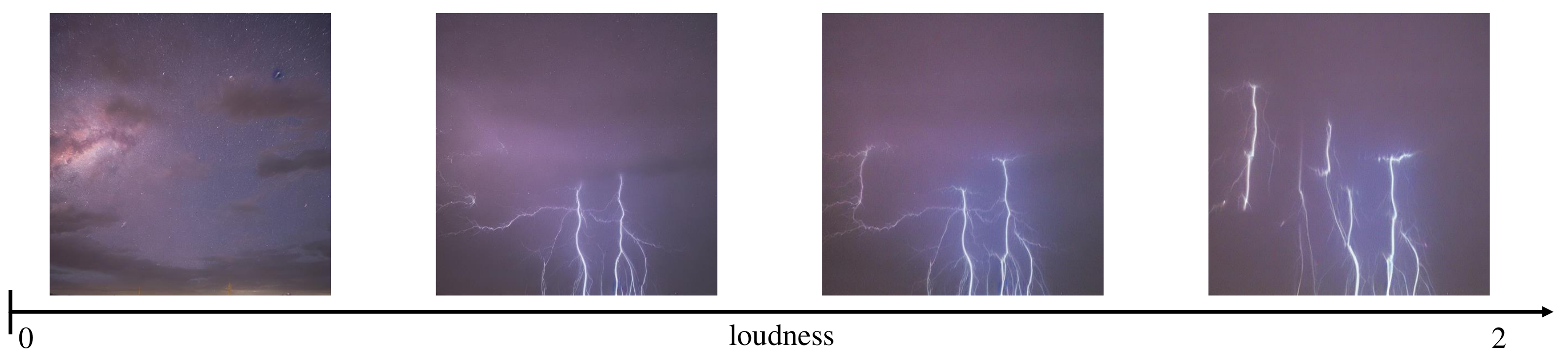}
}
\caption{Thunder intensifies with the loudness strength.}
\vspace{-4pt}
\label{fig:loud}
\end{figure}

Fig.~\ref{fig:loud} presents an example featuring thunder. The progression of thunder loudness is mirrored in the degree of the generated $A^*$. As the thunder intensifies, the depiction of the generated thunder gradually envelops more extensive regions of the sky.

\end{document}